\documentclass[twocolumn,prl,showpacs,superscriptaddress,amsmath,amssymb]{revtex4}
\usepackage{graphicx}

\begin{document}
\title{Multiferroic Domain Dynamics in Strained Strontium Titanate\footnote{In review in \textit{Physical Review Letters}}}

\author{A. Vasudevarao}
\affiliation{Department of Materials Science and Engineering, The Pennsylvania State University, University Park, PA 16802, USA}

\author{A. Kumar}
\affiliation{Department of Materials Science and Engineering, The Pennsylvania State University, University Park, PA 16802, USA}

\author{L. Tian}
\affiliation{Department of Materials Science and Engineering, The Pennsylvania State University, University Park, PA 16802, USA}

\author{J. H. Haeni}
\affiliation{Department of Materials Science and Engineering, The Pennsylvania State University, University Park, PA 16802, USA}

\author{Y. L. Li}
\affiliation{Department of Materials Science and Engineering, The Pennsylvania State University, University Park, PA 16802, USA}
\affiliation{MST-STC, Los Alamos National Lab, Los Alamos, NM 87545, USA}

\author{C-J Eklund}
\affiliation{Department of Physics and Astronomy, Rutgers University, NJ 08854, USA}

\author{Q. X. Jia}
\affiliation{MST-STC, Los Alamos National Lab, Los Alamos, NM 87545, USA}

\author{R. Uecker}
\affiliation{Institute for Crystal Growth, Rudower Chaussee 6, D-12489 Berlin, Germany}

\author{P. Reiche}
\affiliation{Institute for Crystal Growth, Rudower Chaussee 6, D-12489 Berlin, Germany}

\author{K. Rabe}
\affiliation{Department of Physics and Astronomy, Rutgers University, NJ 08854, USA}

\author{L. Q. Chen}
\affiliation{Department of Materials Science and Engineering, The Pennsylvania State University, University Park, PA 16802, USA}

\author{D. G. Schlom}
\affiliation{Department of Materials Science and Engineering, The Pennsylvania State University, University Park, PA 16802, USA}

\author{V. Gopalan}
\email{vgopalan@psu.edu}
\affiliation{Department of Materials Science and Engineering, The Pennsylvania State University, University Park, PA 16802, USA}


\begin{abstract}
Multiferroicity can be \textit{induced} in strontium titanate by applying biaxial strain, resulting in the coexistence of both ferroelectric and antiferrodistortive domains.  The magnitude and sign of the strain imposed on the lattice by design can be used to tune the phase transitions and interactions between these two phenomena. Using optical second harmonic generation, we report a transition from centrosymmetric $4/mmm$ phase to ferroelectric $mm2$, followed by an antiferrodistortive transition to a coupled ferroelastic-ferroelectric $mm2$ phase in a strontium titanate thin film strained in biaxial tension by 0.94\%. The results agree well with theoretical first principles and phase-field predictions.  Direct imaging of domains arising from the ferroelectric phase transition, and its switching under electric fields is demonstrated using piezoelectric force microscopy.  Nonlinear optics combined with phase-field modeling is used to show that the dominant multiferroic domain switching mechanism is through coupled $90^\circ$ ferroelectric-ferroelastic domain wall motion. More broadly, these studies of coexisting ferroelectric (polar) and antiferrodistortive rotation (axial) phenomena could have relevance to multiferroics with coexisting ferroelectric (polar) and magnetic (axial) phenomena.
\end{abstract}


\maketitle

Multiferroic materials with multiple order parameters such as polarization, magnetization, and spontaneous strain lead to the coexistence of two or more of the primary ferroic properties.  As a consequence, multiferroics are attracting significant interest due to the possibility of a rich array of coupled phenomena such as ferroelastic-electric-magnetic, piezo-electric-magnetic, and electro-magneto-optic effects \cite{1,2,3,4,5,6,7,8}. At first glance, strontium titanate would appear an unlikely candidate for a multiferoic. Bulk SrTiO$_3$ is a cubic ($m\bar 3m$) centrosymmetric perovskite at room temperature; it undergoes a nonpolar antiferrodistortive (AFD) phase transformation to a tetragonal point group $4/mmm$ at $\sim$105 K, and exhibits indications of a frustrated ferroelectric (FE) transition at $\sim$20 K, which it never completes \cite{9}.  First principles calculations \cite{10} and thermodynamic analysis \cite{11,12} have suggested however, that external strain can \textit{induce} ferroelectricity. This prediction has been experimentally confirmed recently in a SrTiO$_3$ thin film strained in biaxial tension \cite{13}.  A number of fundamental issues, however, remain to be addressed.  No direct imaging of ferroelectric domains or its switching dynamics has been reported in this material.  The multiferroic nature of strained strontium titanate, i.e., the coexistence of ferroelectric and ferroelastic domains has not received much attention, though it is unique in many respects.  It is \textit{induced} by external strain. Further, unlike other ferroelectric-ferroelastics such as BaTiO$_3$ and PbTiO$_3$ which have a primary order parameter in polarization and secondary order parameter in strain, two \textit{independent primary} order parameters exist:  a \textit{polar} ferroelectric order parameter \textbf{p}, and an axial antiferrodistortive rotation order parameter \textbf{q} \cite{12,14,15,16}. Hence, depending on the magnitude and sign (compressive or tensile) of the applied biaxial strain, the ferroelectric transition temperature can precede or succeed the ferroelastic transition temperature \cite{12}. No first principles theory or experimental determination of the structure and polarization direction has been reported for the multiferroic phase. From a symmetry perspective, the SrTiO$_3$ multiferroic domain structure has striking similarities with, and hence of broader relevance to other multiferroics with coexisting polar and axial phenomena, such as materials that exhibit coexisting ferroelectricity (polar property) and antiferromagnetism (axial property) \cite{3,6,17}.

\begin{figure}[t]
\begin{center}
 \includegraphics[width=0.4 \textwidth]{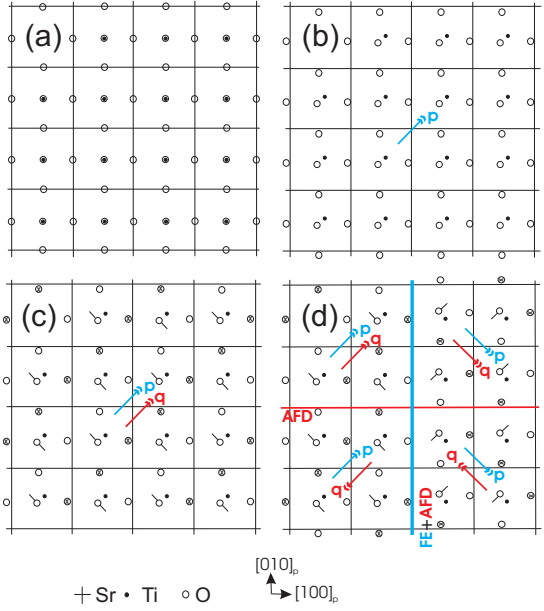}
\end{center}

\caption{\label{fig:1} (Color Online) Schematic showing Sr, O and Ti atomic positions and order parameters \textbf{p} and \textbf{q} in various phases of strained SrTiO$_3$ (a) Paraelectric $4/mmm$ phase, (b) Ferroelectric $mm2$ phase, (c) Ferroelectric (FE) and Antiferrodistortive (AFD) $mm2$ phase. Circles with crosses (dashes) are oxygen ions moving into the plane of the paper (in the dash direction).  (d) Domains in FE+AFD phase showing an AFD and coupled FE+AFD domain wall.}
\end{figure} 

This paper focuses on the domain structure and phase transitions in a 500 \AA~ thick SrTiO$_3$ thin film grown on a (110) DyScO$_3$ substrate with a 0.94\% uniform in-plane biaxial tensile strain in the growth plane \cite{13}.  Thermodynamic calculations predicts a ferroelectric (FE) transition temperature range of $277\pm100$ K, and an antiferrodistortive (AFD) transition at $\sim$111-150 K \cite{12,18}. First principles calculations were performed for this strain value with an $Amm2$ space group symmetry in the FE phase, and an $Ima2$ (\#46) symmetry in the multiferroic (FE+AFD) phase. In the multiferroic phase, the optimized ground state structure has a ferroelectric polarization of \textbf{p} $\sim$0.18 C/m$^2$ in the $\langle110\rangle_p$ direction in the growth plane, and a corresponding antiferrodistortive rotation of the oxygen octahedra by 5.9$^\circ$ about the vector \textbf{q}$\langle110\rangle_p$, where the subscript \textit{p} refers to the pseudocubic Miller index, showing that the combination of the two distortions is lower in energy than either distortion alone.  In this phase, there are 8 types of coexisting FE+AFD domains (see Fig.~\ref{fig:1}(d)), which we classify into 4 types of domain classes: $x+, x-, y+$, and $y-$, each comprised of two types of domains. Denoting the directions $x\equiv[110]_p, y\equiv[1\bar1 0]_p$, and $z\equiv[001]_p$, the order parameters in the 4 types of domain classes are $x+: (p_x, 0, 0, q_x, 0, 0)$ and $(p_x, 0, 0, -q_x, 0, 0)$,  $x-: (-p_x,0, 0, -q_x, 0, 0)$ and $(-p_x,0, 0, q_x, 0, 0)$ ,  $y+:  (p_y, 0, 0, q_y, 0, 0)$ and $(p_y, 0, 0, -q_y, 0, 0)$, and $y-: (-p_y, 0, 0, q_y, 0, 0)$ and $(-p_y, 0, 0, -q_y, 0, 0)$.  These domains are separated by 180$^\circ$ AFD (FE) domain walls across which \textbf{p} (\textbf{q}) remains the same, but \textbf{q} (\textbf{p}) rotates by 180$^\circ$; and 90$^\circ$ (180$^\circ$) FE+AFD domain walls across which both \textbf{p} and \textbf{q} rotate by 90$^\circ$ (180$^\circ$).  In the pure FE phase, the order parameter \textbf{q} disappears, resulting in 4 types of domains, and 2 types (90$^\circ$ and 180$^\circ$) of ferroelectric domain walls.

\begin{figure}[t]
 \begin{center}
 \includegraphics[width=0.5 \textwidth,angle=0]{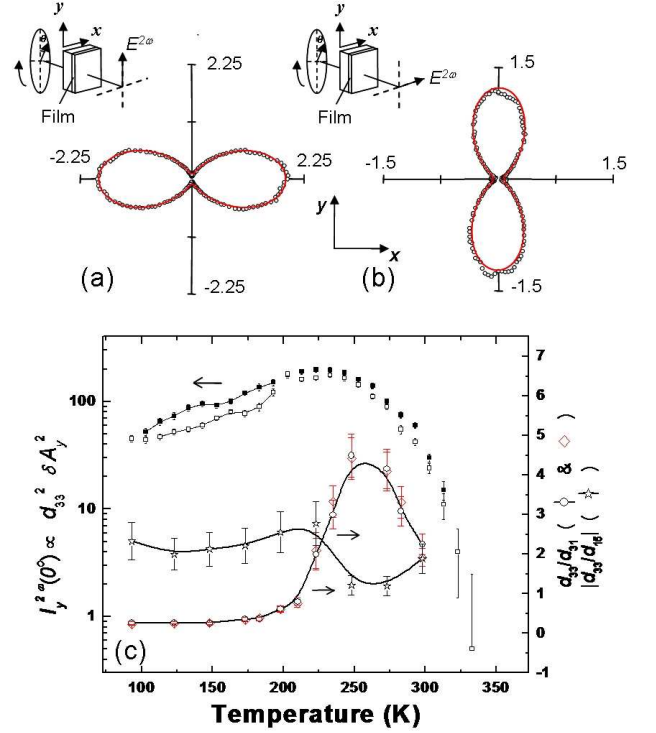}
\end{center}

\caption{\label{fig:2} (Color Online) SHG polar plots at $T=$123 K of (a) $I_y^{2\omega}$ and (b) $I_x^{2\omega}$. (c) $I_y^{2\omega}(\theta=0^\circ)$  vs. $T$ where the dark (hollow) squares are cooling (heating) cycle. The circles (diamonds) represent $K_{3,y}$ ($K_{3,x}^{-1}$).}
\end{figure} 

We employ optical second harmonic generation (SHG) to study domain structures and phase transitions.  SHG involves the conversion of light (electric field $E^\omega$) at a frequency $\omega$ into an optical signal at a frequency $2\omega$ by a nonlinear medium, through the creation of a nonlinear polarization $P_i^{2\omega}\propto d_{ijk} E_j^{\omega} E_k^{\omega}$.  The light source for SHG experiment was a Ti:Sapphire laser with a pulse width of 140 femtoseconds, repetition rate of 1 kHz and  a wavelength of 800 nm. Fundamental wave of 800 nm was incident from the substrate side in normal geometry. Its polarization direction, at an angle $\theta$ from the \textit{y}-axis, was rotated continuously within the plane of the film (see insets in Fig.~\ref{fig:2}(a) and \ref{fig:2}(b)).  The intensity, $I_j^{2\omega}$, of the output SHG signal at 400 nm wavelength from the film was detected along either $j=x, y$ polarization directions by a photomultiplier tube.  The resulting polar plots of the SHG intensity at 123 K are shown in Fig.~\ref{fig:2}.  A reference study of a bare (110) DyScO$_3$ substrate without any film yielded no SHG intensity within the detection limits. Using SHG, the point group of this film at 223 K (ferroelectric phase) was determined to be $mm2$, and the ferroelectric polarization \textbf{p} was found to be along the $\langle110\rangle_p$ axes in the film plane \cite{12}, consistent with first principles calculations. A theoretical analysis of this film system yields the following expression for the SHG intensity \cite{19}, 

\begin{eqnarray}
 \label{eq-1}
I_j^{2\omega} = && K_{1,j} \sin^2 2\theta + K_{2,j} (\sin ^2 \theta + K_{3,j} \cos^2 \theta)^2 \nonumber \\
&&+ K_{4,j}(\sin^2 \theta +K_{3,j}\cos^2 \theta)\sin 2\theta,
\end{eqnarray} 
where $K_{i,j}$ are constants given by:
\begin{equation}
 \label{eq-2}
K_{3,x} = \frac{1}{K_{3,y}} = \frac{d_{31}}{d_{33}},
\end{equation} 

\begin{equation}
 \label{eq-3}
\left( \frac{1}{K_{3,y}} \right)^2 \left( \frac{K_{1,x}K_{1,y}}{K_{2,x}K_{2,y}} \right) = 
\left( \frac{d_{15}}{d_{33}} \right)^4,
\end{equation} 

\begin{equation}
\label{eq-4}
 \left( \frac{K_{1,x}K_{2,y}}{K_{2,x}K_{1,y}} \right)K_{3,y}^2 = \left( \frac{\delta A_y}{\delta A_x} \right)^4.
\end{equation} 
Two of these are ratios of nonlinear coefficients ($d_{31}/d_{33}$ (including its sign) and $d_{15}/d_{33}$ (absolute value only)) that are independent of domain microstructure.  The third quantity, $\delta A_x/\delta A_y$, is a microstructural quantity, where $\delta A_x = A_{x+} - A_{x-}$ and $\delta A_y = A_{y+} - A_{y-}$ are the net area fractional biases in the $x$ and $y$ directions, respectively.  The term $\delta A_x$ ($\delta A_y$) in the observed SHG intensity reflect a cross-cancellation effect of signals from the $x+(y+)$ domains and $x-(y-)$ domains resulting from a phase difference of $\pi$ in their relative SHG fields.  The inverse relation between $K_{3,x}$ and $K_{3,y}$, stipulated by Eq. \ref{eq-2}, can be seen to hold at all temperatures studied (Fig.~\ref{fig:2}(c)), showing that the symmetry is $mm2$ and the polarization is oriented along $\langle110\rangle_p$ directions in the 80-298 K range studied. These properties also show an anomaly in the range of $T_{max}\sim$250-273 K, consistent with the peak in the dielectric response $\varepsilon$ reported in Ref.\cite{13}.  Non-centrosymmetry and polarization, $P_s$, exist well up to \textit{at least} $T_b\sim$323-328 K, where finally the SHG signal is below our detection limit. The SHG signal in Fig. 1(c) also shows a reproducible anomaly at $T_{AFD}\sim148$ K (heating) to $T_{AFD}\sim173$ K (cooling). This corresponds to the antiferrodistortive phase transition in SrTiO$_3$, consistent with thermodynamic phase field predictions \cite{8,9}. The slight drop in the SHG intensity at the AFD transition could arise from  a more effective phase cancellation of the SHG intensity due to doubling of the domain variants from 4 to 8 in the multiferroic phase.  It can also arise from an anomaly in the magnitude of $d_{33}$ across the phase transition, though the measured ratios of nonlinear coefficients in Fig. 2 do not show an anomaly. The symmetry is observed to remain $mm2$ across the AFD transition, as predicted.

\begin{figure}[b]
 \begin{center}
 \includegraphics[width=0.45 \textwidth,angle=0]{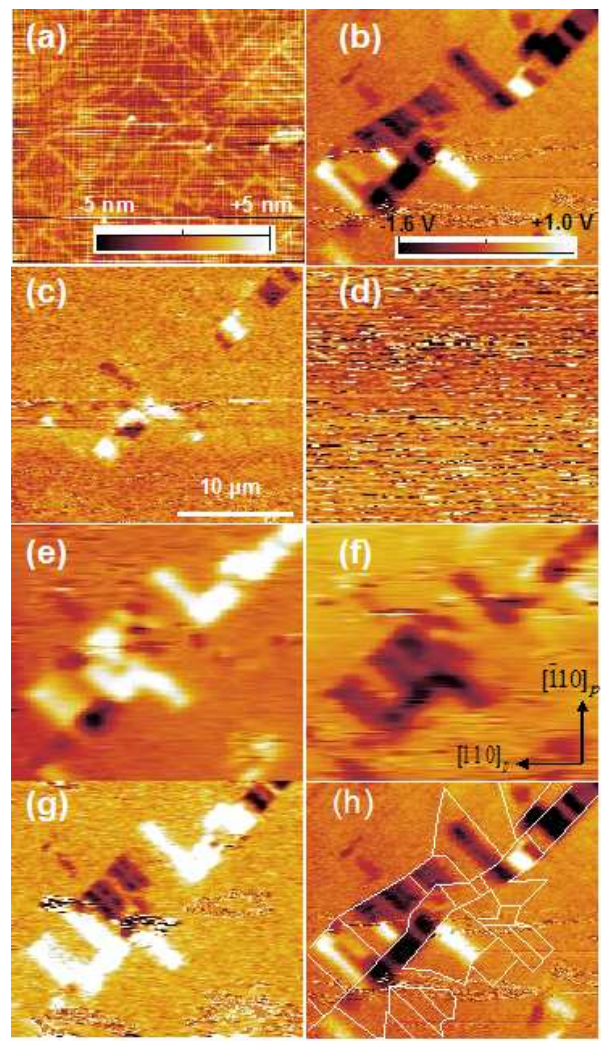}
\end{center}

\caption{\label{fig:3} (Color Online) Topography (a), and lateral PFM images (b-g)  at (b) 298 K (c) 313 K, (d) 333 K, (e) cooled back to 298 K. (f) under field of +1 kV/mm, and (g) -1 kV/mm in the $[110]_p$.  (h) is same as (b) but with overlapped partial topography from (a).}
\end{figure} 

The difference in the temperatures at which the peak in the dielectric constant and the  peak in the $d_{33}/d_{31}$ ratio occurs ($T_{max}\sim$250 K) versus where noncentrosymmetry was lost ($T_b\sim$323 K) can arise from ferroelectric relaxor behavior as recently reported \cite{20}, as well as inhomogeneously strained regions. The $T_b\sim$323-328 K transition may then be more appropriately termed as the Burns temperature \cite{21}. Between $T_{max}$ and $T_b$, polarization in a relaxor ferroelectric is normally expected to exist as glassy nanopolar regions without well-defined domains or domain walls.  This is consistent with piezoelectric force microscopy (PFM) at room temperature (298 K), which reveals that the majority of the film is featureless with low piezoelectric signals. Some localized regions (estimated to be in $\sim$5-10\% of the film area) however reveal ferroelectric domains at room temperature as seen in Fig. 3(b), indicating inhomogeneous strains.  Upon heating, the piezoelectric response starts to disappear partially at 40 $^\circ$C (Fig.~\ref{fig:3}(c)) and completely at 60 $^\circ$C (Fig.~\ref{fig:3}(d)), and the domain features reappear on cooling (Fig. 3(e)) back to room temperature. The domains are \textit{partly} bounded by $\sim$5 nm high surface ridge-like features seen in topography (see Fig. 3(h)), within which they reappear after cooling (Fig. 3(e)). These ridge features have cracks in their center \cite{22} seen typically only in SrTiO$_3$/DyScO$_3$ films thicker than 350 \AA.  Although these films have the narrowest dielectric constant versus temperature peaks reported for SrTiO$_3$ or (Ba,Sr)TiO$_3$ films \cite{20}, the widths are still about three times broader than (Ba,Sr)TiO$_3$ single crystals \cite{23}, hence local inhomogeneous strains and domains can arise. As further evidence, similar films that are 350 \AA~ or less in thickness do not show the topographic ridges, SHG signals or domain like features at room temperature.

The bright and dark regions of PFM signal in Fig.~\ref{fig:3}(b) can be attributed to the $y\pm$ and $x\pm$ domains, respectively. The domain walls (not related to topography; see Fig.~\ref{fig:3}(h)) separating these bright and dark domains are parallel to the $\langle100\rangle_p$ directions, which correspond to 90$^\circ$ domain walls.  These walls can be influenced by external electric fields. Figure~\ref{fig:3}(e) shows that cooling has produced predominantly $y\pm$ domains. When an electric field of +1 kV/mm is applied along the $+x$ direction, they transform to predominantly $x\pm$ domains (Fig.~\ref{fig:3}(f)), and then partially reverse back to $y\pm$ domains after applying -1 kV/mm along the $–x$ direction (Fig.~\ref{fig:3}(g)). These results indicate 90-degree ferroelectric domain wall motion.  This is further confirmed by \textit{in situ} SHG under in-plane electric fields of $\pm$1 kV across a 1.3 mm electrode gap in the $\pm x[110]_p$ direction, which indicates (Figures 4(a) and 4(b)) that both $\delta A_x^2$ and $\delta A_y^2$ change. This indicates that the electric field along the $\pm x$ direction switches domains along the $\pm y$ direction as well, indicating motion of 90$^\circ$ domain walls that couple $x\pm$ domains to the $y\pm$ domains. This mechanism is operative at all temperatures including the multiferroic phase, where SHG indicates the motion of coupled 90$^\circ$ ferroelectric-ferroelastic domain walls.

\begin{figure}[h]
 \begin{center}
 \includegraphics[width=0.4 \textwidth,angle=-90]{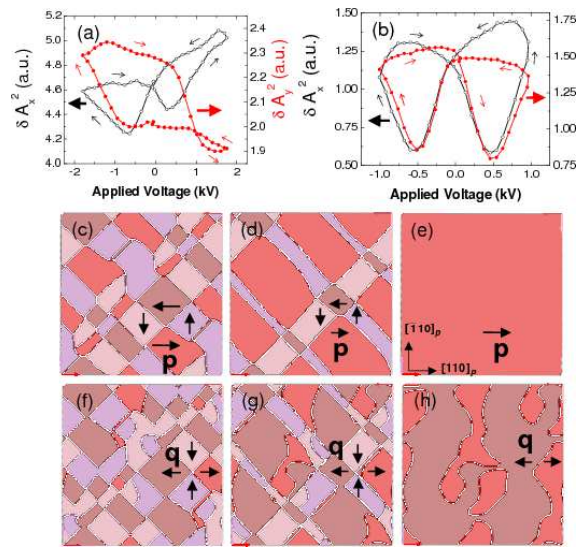}
\end{center}

\caption{\label{fig:4} (Color Online) Domain area fractions ($I_x^{2\omega}(0^\circ)\propto \delta A_x^2$; $I_y^2(0^\circ)\propto \delta A_y^2$)   as a function of applied voltage along the $+x([110]_p)$ direction at (a) $T$=123 K and (b) $T$=223 K.  Phase-field modeling of the evolution of FE ((c)-(e)) and AFD ((f)-(h)) domain morphologies with electric field along $x([110]_p)$.}
\end{figure} 

Phase-field simulations \cite{12,24,25} were performed to understand the detailed mechanism of domain wall motion under an applied electric field  along $+x([110]_p)$ in the multiferroic phase, as shown in Fig.~\ref{fig:4}(c)-(h).  One can clearly see that the area fraction of $x+$ domains increases, while that of $x-$ domains decreases through a motion of coupled 90$^\circ$ ferroelectric-ferroelastic domain walls.  The pure 180$^\circ$ AFD walls (see Fig.~\ref{fig:1}(d)) within a single polarization domain state are degenerate in energy even under an electric field and remain with statistically equal populations (Fig.~\ref{fig:4}(f-h)). The pure AFD domain walls within the polarization domain states not favored by the external field disappear, however, primarily as the 90$^\circ$ degree walls sweep through that domain state.

In conclusion, a unique form of multiferroicity, namely an \textit{induced} ferroelectric-antiferrodistortive phase in strained SrTiO$_3$ is reported. A ferroelectric transition temperature $T_{max}\sim$250 K, Burns temperature $T_b\sim$323 K, and an antiferrodistortive transition at $T_{AFD}\sim$160 K have been determined using optical second harmonic generation (SHG).  For the first time, ferroelectric domains were directly imaged at room temperature and above in this material using PFM. \textit{In situ} SHG under external fields show evidence for the movement of coupled ferroelectric-ferroelastic 90$^\circ$ domain walls in the multiferroic phase under external fields, in agreement with phase-field modeling.  From a symmetry perspective, the domain structure and dynamics in this ferroelectric-antiferrodistortive system could have relevance to expected domain structures in a ferroelectric-antiferromagnetic system with two phase transitions.

\begin{acknowledgments}
We thank C. Fennie for his help in performing first principle calculations. The financial support from NSF under Grant Nos. DMR-0122638, DMR-0507146, DMR-0512165, DMR-0349632, and NSF-MRSEC center at Penn State is gratefully acknowledged.  Chen would also like to acknowledge support from Guggenheim fellowship and Los Alamos National Labs.
\end{acknowledgments}

\end{document}